\documentclass[a4paper]{article}

\usepackage{INTERSPEECH_v2}
\usepackage{amssymb}
\usepackage{graphicx}
\usepackage{amsmath,graphicx}
\usepackage{tikz}
\usepackage{blindtext}
\usepackage{graphicx}
\usepackage{tikz}
\usetikzlibrary{arrows.meta}
\usetikzlibrary{shapes,decorations,arrows,shadows,shapes.geometric}
\usepackage[colorlinks]{hyperref}
\usepackage[colorinlistoftodos]{todonotes}
\usepackage{smartdiagram}
\usetikzlibrary{chains}
\usepackage{amsmath,bm,times}
\usepackage{amsmath}
\usetikzlibrary{mindmap,trees}
\usetikzlibrary{trees}
\usepackage{mathtools}   

\newcommand{\argmax}{\operatornamewithlimits{argmax}}
\usepackage{algorithm}
\usepackage{algorithmic}
\usepackage{tikzscale}
\usepackage{filecontents} 
\usepackage{blindtext}
\usepackage{graphicx}
\usepackage{tikz}
\usetikzlibrary{shapes,snakes}
\usetikzlibrary{shapes,arrows}
\usepackage[colorlinks]{hyperref}
\usepackage[colorinlistoftodos]{todonotes}
\usepackage{smartdiagram}
\usetikzlibrary{chains}
\usetikzlibrary{shapes,arrows,shadows}
\usepackage{amsmath,bm,times}
\usepackage{multirow}

\title{Towards a Knowledge Graph based Speech Interface}
\name{Ashwini Jaya Kumar$^1$, S\"oren Auer$^{1,2}$, Christoph Schmidt$^1$, Joachim K\"ohler$^1$, }
\address{
  $^1$Fraunhofer IAIS, Schloss Birlinghoven, Sankt Augustin, Germany\\
  $^2$University of Bonn, Bonn, Germany}
\email{\footnotesize \{ashwini.jaya.kumar,soeren.auer,christoph.andreas.schmidt,joachim.koehler\}@iais.fraunhofer.de}

\begin{document}

\maketitle
\begin{abstract}

Applications which use human speech as an input require a speech interface with high recognition accuracy. The words or phrases in the recognised text are annotated with a machine-understandable meaning and linked to knowledge graphs for further processing by the target application. %These semantic annotations of recognised words can be represented as a subject-predicate-object triples which collectively form a graph often referred to as a knowledge graph. 
This type of knowledge representation facilitates to use speech interfaces with any spoken input application, since the information is represented in logical, semantic form., retrieving and storing can be followed using any web standard query languages. 
In this work, we develop a methodology for linking speech input to knowledge graphs and study the impact of recognition errors in the overall process. 
We show that for a corpus with lower WER, the annotation and linking of entities to the DBpedia knowledge graph is considerable.
DBpedia Spotlight, a tool to interlink text documents with the linked open data is used to link the speech recognition output to the DBpedia knowledge graph. 
Such a knowledge-based speech recognition interface is useful for applications such as question answering or spoken dialog systems.
\end{abstract}


\section{Introduction}

A speech interface to any application involves speech recognition, transcript interpretation and semantic representation useful for the target application.  
Speech recognition is the process of recognising the spoken utterance and generating the transcription of the audio input. 
The transcription has to be interpreted to understand the meaning of the sentence. 
Interpreting a sentence involves parsing a sentence and representing its linguistic constituents using a machine-processable semantic knowledge representation formalism. 
%This representation of the linguistic constituents is referred as semantic representation. 
While there are a number of knowledge representation formalisms (e.g. frames, topic maps), the currently prevailing semantic representation employs atomic facts (or statements) represented as Resource Description Framework (RDF) triples.
RDF triples consist of subject, predicate and object, where entities at subject or object position are connected via properties at the predicate position (all identified using URI/IRIs). 
This type of semantic representation makes it easy for the target application to derive the meaning in the recognised transcripts and facilitates the integration of the speech interface component into the target application. 

Based on RDF, the Linked Data paradigm~\cite{DBLP:conf/rweb/AuerLNZ13} aims at connecting related data on the Web by reusing and referring to existing data and schema elements using the respective IRI/URI identifiers. 
Interlinking the output of a speech recogniser to Linked Open Data (LOD) has the advantage of leveraging the comprehensive background knowledge meanwhile available on the Web of Data. 
The Linked Data paradigm provides an easy mode to share, re-use and interlink data. 
This helps a speech interface component to be dynamically used for various applications with human speech as input. 
Linking text to LOD increases the discoverability of information~\cite{Mendes:2011} in the context of the spoken utterance. 
%The class of each node in LOD helps to automatically annotate the recognised text to it's classes.

To the best of our knowledge there is no prior work on directly leveraging knowledge graphs for speech recognition.
However, knowledge graphs were, for example, used for spoken response and for language modeling.
For spoken response, \cite{de2015dialogue} shows how Linked Data can be used for a voice-based mobile phone application. 
%A market information system called RadioMarche is developed and the obtained market data is exposed as Linked Data. 
%A voice output response using the open standardised \emph{VoiceXML} language is provided by assembling snippets of pre-recorded audio files linked to the users query. 
Using a knowledge graph for natural language processing, in particular, for language modeling is introduced in \cite{ahn2016neural}. 
%It is often difficult to include all the available words in a particular language to the vocabulary and also depending on a huge corpus is infeasible in the present web of data. 
%The conventional language model depends on statical co-occurrences of words and fails to deal with unknown words. 
The authors propose a method which uses factual knowledge in the knowledge graph and is able to predict unknown entities. 
To the best of our knowledge, in existing speech interface systems the use of knowledge graphs for semantic grounding of speech recognition is not explored. 
However, some initial recent efforts to use a knowledge graph for language understanding are presented in~\cite{hakkani2013using, akbacak2014rapidly, heck2012exploiting}.

In this article, we are extending the idea of using a knowledge graph for language understanding to speech interfaces and study the effects of speech recognition errors on the linking process. 
We recognise the spoken utterance followed by entity annotation and entity linking to the knowledge graph.
As a result the grounding of the recognised text in the knowledge graph can be exploited for error detection and applications such as voice-based semantic search, question answering or spoken-dialog systems.

The main \textbf{contributions} of this work are:
\begin{enumerate}
    \item{A novel approach for a knowledge-graph-based speech interface which can be applied to applications with speech input, e.g. Question Answering and Spoken Dialog Systems}
    \item Demonstration of the use of a knowledge graph for automatic speech recognition and study of the effects of recognition errors on the knowledge graph interface.
\end{enumerate}

Also, with this work we do a first step towards the integration of the two still largely independent research domains: Semantic Web and Automatic Speech Recognition.

\begin{figure}[tb]
    \centering
    \resizebox{0.5\textwidth}{!}{\includegraphics{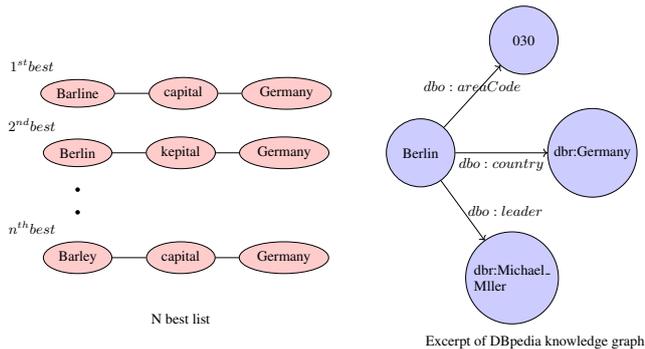}}
    \caption{N-best list for the sentence `Berlin is the capital of Germany' (left) and excerpt of the corresponding entity description for Berlin in the DBpedia knowledge graph (right).}
\label{fig:nbest}
\end{figure}

\section{Motivation}

Current speech recognition systems still produce output which contains recognition errors. 
It is very difficult to train a recogniser for all the entities in a particular language. 
Recognising entities is an important factor in applications such as question answering and spoken dialog systems. 
Most of the time, the recogniser is built in a way such that an unrecognised word is substituted with a similar sounding word within a given context. 
Given two entities, predicting a third entity is straightforward if the combination of those three entities constitutes a triple in a knowledge graph. 
Consider a sentence with three entities, if a wrong entity is substituted in the speech recognition process, replacing that with the more appropriate one available in the knowledge graph intuitively reduces the recognition errors. 
A connection in the underlying knowledge graph between entities linked to words in a sentence provides evidence about the logical connection of these entities and thus about the probability of relatedness between the words.
In addition, the entity description in the knowledge graph provides information about further entity attributes. 
A systematic analysis and usage of links to an underlying knowledge graph and connections within the graph can result in improved recognition accuracy. 

For example: Consider the sentence  ``Berlin is the capital of Germany.'', which can be represented in logical triple form as (\texttt{<Berlin, capital, Germany>}).  
The entities referred to in the above sentence are \texttt{Berlin} and \texttt{Germany} both connected through the property \texttt{capital}. 
Figure~\ref{fig:nbest} shows how the connection information in the knowledge graph can be related to the words annotated in the N-best list. 
There is an erroneous substitution in the $1^{st}$ best list of the sentence, instead of \texttt{Berlin}, \texttt{Barline} word is substituted. 
If the other two words are correctly recognised and linked to respective entities, then by considering the knowledge graph triple above connecting \texttt{Berlin} and \texttt{Germany}, it is possible to identify the error and replace the wrong substitution with the correct one.

% Define the layers to draw the diagram
\pgfdeclarelayer{background}
\pgfdeclarelayer{foreground}
\pgfsetlayers{background,main,foreground}

% Define block styles used later

\tikzstyle{sensor}=[draw, fill=blue!20, text width=5em, 
    text centered, minimum height=2em]
\tikzstyle{ann} = [above, text width=5em, text centered]
\tikzstyle{wa} = [sensor, text width=8em, fill=red!20, 
    minimum height=4em, rounded corners]
\tikzstyle{sc} = [sensor, text width=10em, fill=red!20, 
    minimum height=7em, rounded corners]

% Define distances for bordering
\def\blockdist{2.3}
\def\edgedist{2.5}

\begin{figure}[tb]
\centering
\resizebox{0.50\textwidth}{!}{\includegraphics{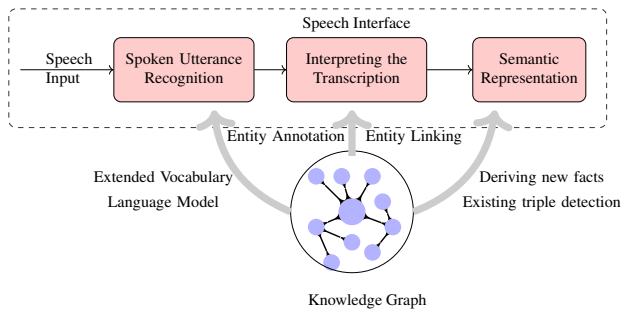}}
\caption{Using a Knowledge Graph in various Speech Interface stages.}
\label{fig:slrucom}
\end{figure}

\section{Linking Speech to Knowledge Graphs}

A speech interface to applications with speech as input involves recognising and interpreting the spoken utterance. 
In the proposed approach for a speech interface, we are connecting entity references in the output of a speech recogniser to entity descriptions represented as RDF triples contained in a triple store. 
After transcribing the audio input, entity annotation and entity linking are performed in the process of grounding the speech transcription in the knowledge graph. 
The use of a knowledge graph in various stages of a speech interface is shown in Figure~\ref{fig:slrucom}.
In the speech recognition phase, the entity descriptions contained in the knowledge graph can be used to extend the vocabulary with domain-specific terms or provide a fact-based language model.
During the interpretation of the transcript, entity references in the text are linked to corresponding entity descriptions and typed with respective classes in the knowledge graph.
Finally, the semantic representation can either involve the detection of existing triples in the knowledge graph representing the meaning of the spoken utterance or discovering new triples by extracting relations between entity references.

\tikzstyle{block1} = [draw, fill=blue!20, rectangle, align=center,
    minimum height=4.0em, minimum width=19em, rounded corners]
\tikzstyle{block2} = [draw, fill=white!20, rectangle, minimum height=2em, minimum width=18em, rounded corners]
\tikzstyle{input} = [coordinate]
\tikzstyle{output} = [coordinate]
\tikzstyle{pinstyle} = [pin edge={to-,thin,black}]

\begin{figure}[tb]
\centering
\resizebox{0.3\textwidth}{!}{\includegraphics{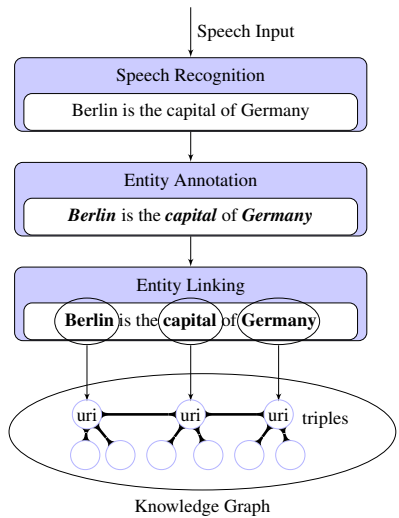}}
\caption{Speech input linking steps}
\label{fig:link}
\end{figure}

\textbf{Linking Framework.}
Linking speech input to a knowledge graph involves recognising the spoken input, identifying the words in the recognised sentence which are referring to entities in the knowledge graph and linking them to the corresponding entity descriptions in the knowledge graph. The linking framework is described in Figure~\ref{fig:link}.

 Speech recognition can be described as a function mapping from a sequence of observations into a sequence of words. The sequence of observations are the features extracted from the audio signal. Let $O=\{o_1, o_2,\dots,o_M\}$ be a set of observation sequences and $W=\{w_1, w_2,\dots,w_N\}$ be a set of words. The probability of recognising the word sequence given the observation sequence can be written as,

\begin{equation}
\hat{W} = \underset{}{\argmax_W} P(W|O)
\label{eq:as}
\end{equation}

$\hat{W}$ signifies the word with highest probability for a given observation sequence. 
However, $P(W|O)$ is difficult to obtain directly and hence we can use the Bayes' rule,

\begin{equation}
P(W|O) = \frac{P(O|W)P(W)}{P(O)}
\label{eq:bayes}
\end{equation}

In Equation~\ref{eq:bayes}, $P(W|O)$ is the a posteriori probability of predicting the word sequence given the observation sequence, $P(O|W)$ is the likelihood probability (i.e. acoustic model) and $P(W)$ is the a priori probability of the word sequences (i.e. language model). 
The term $P(O)$ can be neglected since it remains constant.
Hence Equation~\ref{eq:as} becomes,

\begin{equation}
\hat{W} = \underset{}{\argmax_W}P(O|W)P(W)
\label{eq:asrf}
\end{equation}

A knowledge graph comprises (subject, predicate, object) triples which connects nodes or vertexes (i.e. subject and object) via edges (predicate). 
A statement is represented as a triple as $<e_i,r,e_j>$, where $e_i$ and $e_j$ are entities (e.g. a Person or an Actor) and $r$ is the relation between both (e.g. spouse\_of or played\_in\_movie). Such statements follow the Resource Description Framework (RDF), which is a W3C standard for knowledge representation. 
Based on these triples, an RDF graph can be built where each entity (e.g. a Person) is described by a number of triple statements having the entity as subject and linking it to other entity descriptions (e.g. relatives or movies). 
In addition, attributes can be used to attach further information as literal values (such as text labels, birth dates) to the entities. 

One of the most commonly used knowledge graphs is DBpedia~\cite{DBLP:journals/semweb/LehmannIJJKMHMK15}, which represents statements extracted from Wikipedia.
It comprises more than 1 billion triples describing more than 20 million entities covering all domains present in Wikipedia (e.g. people, organisations, places, movies, diseases etc.). An excerpt of DBpedia knowledge graph is shown in the right portion of Figure~\ref{fig:nbest}.

The words in the recognised sentence which are also entities in the knowledge graph have to be identified. 
This can be done using a simple string matching algorithm between recognised words and entity labels. If there is no exact match, the linking algorithm is expected to find the most relevant entity label.
We are framing the problem of entity annotation as a probabilistic model.
Entity annotation inherently assumes that both entity annotation and entity linking is done simultaneously since the annotated entities are linked to a knowledge graph. 
Let $E=\{e_1, e_2,\dots,e_K\}$ be the set of entity resources in a knowledge graph. 
Then the probability of linking the word sequence to the entities is the probability of finding the most relevant entity in the knowledge graph entity resources which can be written as $P(E|W)$. The probability of predicting an entity resource for a given word can be written as,

\begin{equation}
\hat{E} = \underset{}{\argmax_E} P(E|W)
\label{eq:ent}
\end{equation}
Deducing the above equation using Bayes' rule as in Equation~\ref{eq:bayes} and Equation~\ref{eq:asrf} we obtain,

\begin{equation}
\hat{E} = \underset{}{\argmax_E}P(E)P(W|E)
\label{eq:ent}
\end{equation}

The likelihood probability $P(W|E)$ can be obtained by any statistical modeling methods and $P(E)$ is the a priori probability of the entity resource. Apart from probabilistic approach, there are several other ways of matching a word sequence to entity resource. Exact match models and vector space approaches are a few to mention.

\section{Implementation}\label{sec:imp}

The entity annotation has to be performed in such a way that textual entity references are connected to the triples in a knowledge graph.
We are using \textit{DBpedia Spotlight}~\cite{Mendes:2011} for that purpose.
DBpedia Spotlight is a tool for interlinking text documents with Knowledge Graphs which enables to use the knowledge embedded in the documents for relevant applications. 
It automatically annotates the text documents and connects those to DBpedia entities (identified via URIs). 
Details on spotting the phrase in a sentence, candidate selection, disambiguation and configuration are presented in~\cite{Mendes:2011}.

DBpedia Spotlight computes scores such as prominence, topic pertinence and contextual ambiguity. 
Prominence computes how many times a resource is mentioned in Wikipedia. 
%A term called anchor is used to refer to a phrase in Wikipedia articles to which the links are attached. 
%In calculating prominence, the number of such anchors is considered. 
%Spotlight selects all the entities that are greater than the informed value (which by default is 10). 
A term called support is used for prominence and priorscore for normalised prominence score in DBpedia Spotlight software. 
Topic pertinence measures how relevant the annotated resource is to the DBpedia resource. 
It uses the similarity score from the disambiguation step to compute the relevance. 
It is called percentageOfSecondRank in the DBpedia Spotlight software. 
If the topical relevance to a paragraph has more than one candidate, then it might be hard to disambiguate. 
The relative difference of the topical relevance between the first and second candidate resource gives the contextual ambiguity score (contextual\_score in DBpedia Spotlight) which helps to know by how much the first and second candidate resources are confused. 
A combination of all these scores is called finalScore.

\section{Experimental Evaluation}

\begin{table*}[tb]
\begin{center} \footnotesize
\begin{tabular}{ |l|c|r|r|r|r|r|r|r } 
\hline
\textbf{Database} & \textbf{Test set labels} & \textbf{WER (\%)} & \textbf{Test sentences} & \textbf{Entities (Ref)} & \textbf{Entities (Test)} & \textbf{Difference (\%)}\\
\hline
\multirow{2}{4em}{\textit{WSJ}} & Eval92 & 5.64 & 333 & 1,802 & 1,836 & 1.02\\ 
& Eval93 & 9.02 & 213 & 1,129 & 1,139 & 1.00 \\ 
\hline
\textit{Tedlium} & Test set & 31.85 & 1,155 & 5,804 & 6,743 & 1.16\\
\hline
\end{tabular}
\end{center}
\caption{ASR performance and entity annotation for reference and test sentences in the WSJ and the TED-LIUM corpus}
\label{wer:aa}
\end{table*}

We are using the \textit{Wall Street Journal} (LDC93S6A and LDC94S13A) and the \textit{TED-LIUM} corpus for training and evaluation of the ASR performance, and the \textit{Kaldi}~\cite{povey2011kaldi} toolkit for development. 
The Wall Street Journal (WSJ) corpus mainly consists of broadcast news read by journalists and recorded using a single channel microphone with a sampling frequency of 16kHz. 
A sample sentence looks like ``In Tokyo foreign exchange trading yesterday the yen increased against the dollar''. 
The TED-LIUM corpus was created from TED talks and their transcripts on the website~\cite{rousseau2014enhancing} with a sampling frequency of 16kHz. 
A TED-LIUM sample sentence looks like ``i'd like to share with you a discovery that i made a few months ago while writing an article for italian wired i always keep my thesaurus handy whenever i'm writing anything but''.

Details on the number of sentences used for testing, the ASR performance and the number of entities annotated in the reference and test sentences are presented in Table~\ref{wer:aa}. 
The WSJ test set have shown better recognition performance compared to TED-LIUM. 
The reference and the test sentences are annotated using DBpedia Spotlight. 
As can be seen in Table~\ref{wer:aa}, the percentage difference in the number of entities between reference and test sentences is relatively small in the case of WSJ compared to TED-LIUM. 
The number of entities is higher for test sentences since it also contains inserted and substituted words in place of Out-Of-Vocabulary (OOV) words.

\begin{figure}[tb]
    \centering
    \resizebox{\columnwidth}{!}{\includegraphics{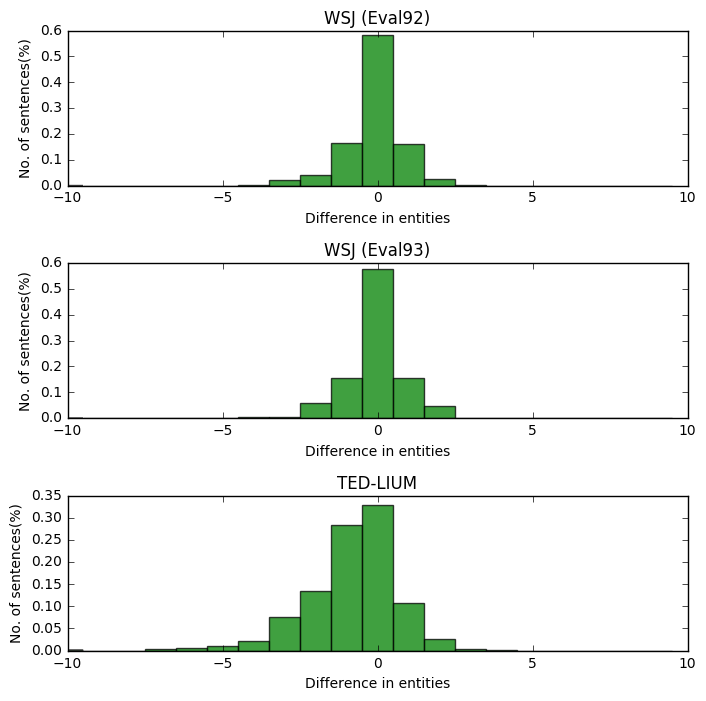}}
    \caption{Histogram of difference in no. of entities per sentence }
\label{fig:f1}
\end{figure}

\begin{figure}[tb]
    \centering
    \resizebox{\columnwidth}{!}{\includegraphics{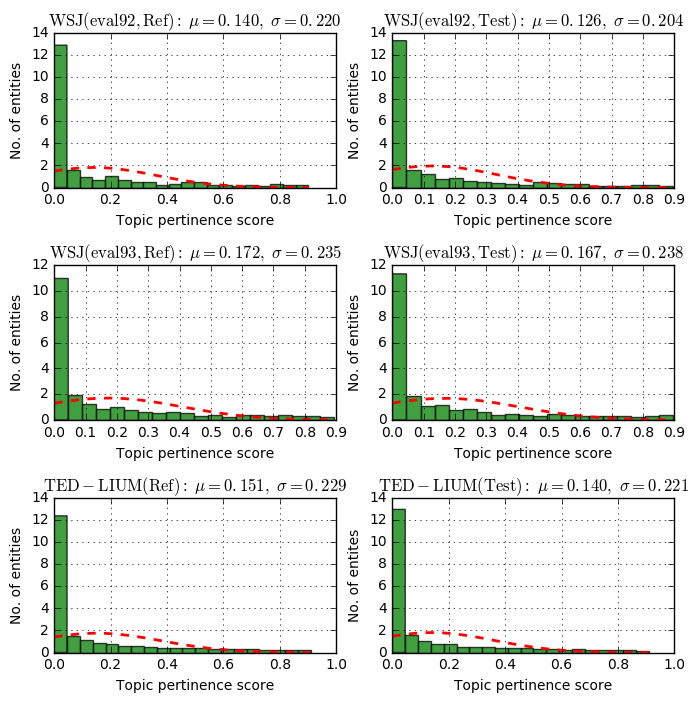}}
    \caption{Histogram of topic pertinence score obtained from DBpedia Spotlight }
\label{fig:f2}
\end{figure}

The distribution of the difference in the number of entities per sentence is shown in Figure~\ref{fig:f1}. 
The difference value is computed by subtracting the number of entities in reference sentence with the number of entities in the test sentence.
If the difference value is positive then the number of entities annotated in the reference sentence is larger than in the test sentence which is due to the missing words in the test sentence. 
Vice versa, if the difference value is negative then the number of entities annotated in the test sentence is larger than in the reference sentence which is due to the inserted and the substituted words in the test sentence. 
In case of the WSJ (eval92 and eval93), the difference value is equally distributed on both positive and negative axes. 
Some entities recognised in the reference sentences were not recognised in the test sentences and vice versa.
In case of TED-LIUM, the difference value is more shifted towards the negative axis, which in turn means that a higher number of extra entities are annotated per sentence for test sentences.
This type of error is typically due to the Out-Of-Vocabulary word problem in ASR. 
The DBpedia Spotlight performance contributes minimally to the difference in number of entities between reference and test sentences. 
Since for a default parameter setting, it is most likely that the DBpedia Spotlight performance remains same for any given text input. 
Even though the test sentences contain errors compared to the reference sentences, DBpedia Spotlight treats both test and reference sentences as text.

The topic pertinence parameter (cf. Section~\ref{sec:imp}) in DBpedia Spotlight gives the relevance score for the annotated word in a sentence wrt. to the DBpedia entity. 
If the topic pertinence value is higher then the word is more likely to refer to the DBpedia entity.  
The distribution of the topic pertinence score for WSJ and TED-LIUM corpus along with mean and variance values are shown in Figure~\ref{fig:f2}. 
If the topic pertinence value is closer to 1.0 then the relevance to a DBpedia resource is stronger. The mean value for reference sentences is greater than for the test sentences which hints about how relevant the annotated words in the reference and test sentences is to the DBpedia entries.

\section{Related Work}

In \cite{raimond2012automated}, a framework to tag the BBC programme archive using DBpedia as a source of tag identifiers is proposed. 
This is done using speech recognition, text processing and concept tagging techniques. 
These techniques have shown positive signs for bootstraping the interlinking process of the archived content. 
In~\cite{gesmundo2014projecting}, the information presented in the knowledge graph is used to improve the performance of a statistical dependency parser. 
The focus of the work is on recognition of relations such as coordination and apposition. 
Apposition is a relation between two adjacent noun phrases and coordination  between nouns relates two or more elements of the same kind. 
In~\cite{44316} the information about entities encoded in FreeBase notable types is used for relation extraction. 
FreeBase notable types are simple atomic labels given to entities that indicate  what the entity is notable for, and so serve as a useful information source. 
\cite{weston2013connecting} uses both labeled text data and triples from the knowledge base for relation extraction. 
%The relation mentions, entities and relationships are all embedded into a common low-dimensional vector space and used to train the model.

\section{Conclusion}

A speech interface to applications with speech as input involves recognising and interpreting the spoken utterance. A knowledge graph based speech interface system which can be used for any application with speech input is discussed. We presented a methodology of linking the speech input to a knowledge graph. Speech recognition, which constitutes the first step of linking, is prone to show errors and its performance is significant for the success of the overall performance of the application it is designed for. In this work, we evaluated the impact of recognition errors on the linking process of speech input to a knowledge graph. We see this work as the first step in a larger research agenda. In particular exploring the use of semantic information from knowledge graphs to improve the recognition accuracy is an extremly promising research direction.

\section{Acknowledgements}
Parts of this work received funding from the European Union’s Horizon 2020 research and innovation program under the Marie Sklodowska-Curie grant agreement No. 642795 (WDAqua project).

\bibliographystyle{IEEEtran}

\bibliography{main.bbl}

% \begin{thebibliography}{9}
% \bibitem[1]{Davis80-COP}
%   S.\ B.\ Davis and P.\ Mermelstein,
%   ``Comparison of parametric representation for monosyllabic word recognition in continuously spoken sentences,''
%   \textit{IEEE Transactions on Acoustics, Speech and Signal Processing}, vol.~28, no.~4, pp.~357--366, 1980.
% \bibitem[2]{Rabiner89-ATO}
%   L.\ R.\ Rabiner,
%   ``A tutorial on hidden Markov models and selected applications in speech recognition,''
%   \textit{Proceedings of the IEEE}, vol.~77, no.~2, pp.~257-286, 1989.
% \bibitem[3]{Hastie09-TEO}
%   T.\ Hastie, R.\ Tibshirani, and J.\ Friedman,
%   \textit{The Elements of Statistical Learning -- Data Mining, Inference, and Prediction}.
%   New York: Springer, 2009.
% \bibitem[4]{YourName17-XXX}
%   F.\ Lastname1, F.\ Lastname2, and F.\ Lastname3,
%   ``Title of your INTERSPEECH 2017 publication,''
%   in \textit{Interspeech 2017 -- 18\textsuperscript{th} Annual Conference of the International Speech Communication Association, August 20?24, Stockholm, Sweden, Proceedings, Proceedings}, 2017, pp.~100--104.
% \end{thebibliography}

\end{document}